\documentclass{llncs}
\usepackage{hyperref}

\usepackage{subfigure}

\usepackage{multirow}

\usepackage[]{graphicx}

\usepackage{newclude}

\usepackage{array}

\usepackage{acro}

\usepackage{todonotes}

\usepackage{booktabs}

\usepackage{amssymb, amsmath}
\usepackage{pifont,amsfonts} 

\usepackage{cleveref}

\usepackage{siunitx}
\DeclareSIUnit\px{px}

\usepackage{fancyhdr}

\begin{document}

\title{Pay Attention with Focus: A Novel Learning Scheme for
  Classification of Whole Slide Images}
\titlerunning{Pay Attention with Focus: FocAtt-MIL}

\author{Shivam Kalra\inst{1,2} \and
Mohammed Adnan\inst{1,2} \and
Sobhan Hemati\inst{1} \and
Taher Dehkharghanian\inst{3} \and
Shahryar Rahnamayan \inst{3} \and
H.R. Tizhoosh\inst{1,2}
}
\authorrunning{Kalra et al.}
\institute{Kimia Lab, University of Waterloo, Waterloo, Canada
  \and
  Vector Institute, MaRS Centre, Toronto, Canada
  \and
  NICI Lab, Ontario Tech University, Oshawa, Canada
}

\pagestyle{fancy}
\lhead{Accepted in MICCAI, 2021}
\rhead{}

%

\maketitle              

\begin{abstract} 
Deep learning methods such as convolutional neural networks (CNNs) are difficult
to directly utilize to analyze whole slide images (WSIs) due to the large image dimensions. We overcome this limitation by proposing a novel two-stage approach. First, we extract a set of representative patches (called mosaic) from a WSI. Each patch of a mosaic is encoded to a feature vector using a deep network. The feature extractor model is fine-tuned using hierarchical target labels of WSIs, i.e., anatomic site and primary diagnosis. In the second stage, a set of encoded patch-level features from a WSI is used to compute the primary diagnosis probability through the proposed \emph{Pay Attention with Focus} scheme, an attention-weighted averaging of predicted probabilities for all patches of a mosaic modulated by a trainable focal factor. Experimental results show that the proposed model can be robust, and effective for the classification of WSIs.
\end{abstract}

\section{Introduction} 
The success of deep learning has opened promising horizons for digital
pathology. AI experts and pathologists are now working together to design novel
image analysis algorithms. The last decade has witnessed the widespread adoption
of digital pathology, leading to the emergence of machine learning (ML) models
for analyzing whole slide images (WSIs). The major applications of ML in digital
pathology include (i)~reducing the workload on pathologists, and (ii)~improving
cancer treatment procedures~\cite{madabhushi2016image}. The computational
analysis of WSIs offers various challenges in terms of image size and
complexity. These challenges necessitate the inquiry into more effective ways of
analyzing WSIs. CNNs are at the forefront of computer vision, showcasing
significant improvements over conventional methodologies for visual
understanding~\cite{khan2020survey}. However, CNNs can not be directly utilized
for processing WSIs due to their large image dimensions. The majority of the
recent work analyzes WSIs at the patch level that requires manual delineations
from experts. These manual delineations reduce the feasibility of such
approaches for real-world scenarios. Moreover, most of the time, labels are
available for an entire WSI and not for individual
patches~\cite{adnan2020representation}. Therefore, to learn a WSI
representation, it is necessary to leverage the information present in all
patches. Hence, multiple instance learning (MIL) is a promising venue for
vision-related tasks for WSIs.

The paper's contribution is three-fold (i) we propose a novel attention-based MIL
approach for the classification of WSIs, (ii) we fine-tune a feature extractor model using multiple and hierarchically arranged target labels of WSIs, and (iii) we present insights of the model's
decision making by visualizing attention values. The method is tested on two
large-scale datasets derived from The Cancer Genomic Atlas (TCGA) repository
provided by NIH~\cite{tomczak2015cancer}.

\section{Background} \label{sec:relwork}
CNN based methods for analyzing histopathological images is well represented in
the literature~\cite{dimitriou2019deep, coudray2018classification,
mahmood2020artificial, gao2018sd}. Deep learning methods generalize well across
patients, disease conditions, and are robust to the vendor or human-induced
variations, especially when a large amount of training
data is available~\cite{dimitriou2019deep}.

A WSI usually contains at least two target labels, anatomic site, and primary
diagnosis that are arranged in a hierarchy. The
simplest way to deal with multi-label classification with $k$ labels is to treat
this as $k$ independent binary classification. Although this approach may be
helpful, it does not capture label dependencies. This limitation can degrade the
performance in many applications where there is strong dependency among labels,
for example, in WSI classification. To address this limitation, two different
approaches, i.e., transformation and algorithm adaption methods, have been
proposed~\cite{zhang2013review}. In transformation-based methods, multi-label
data is converted to new single label data to apply regular single-label
classification. On the other hand, in the adaptation-based category, this is
attempted to modify the basic single-label algorithm to handle multi-label
data~\cite{tidake2018multi}.

There are two main methods for characterizing WSIs~\cite{barker2016automated}.
The first method is called sub-setting, which considers a small section of a
large WSI as an essential region for analysis. On the other hand, the tiling
method, segments a WSI into smaller and controllable patches (i.e.,
tiles)~\cite{gutman2013cancer}. The tiling or patch-based methods can benefit
from MIL. Isle et al. used MIL for digital pathology and introduces a different
variety of MIL pooling functions~\cite{ilse2020deep}. Sudarshan et al. used MIL
for histopathological breast cancer image
classification~\cite{sudharshan2019multiple}. Permutation invariant operator for
MIL was introduced by Tomczak et al. for WSIs processing~\cite{tomczak2017deep}.
Graph neural networks (GNNs) have also been used for MIL applications because of
their permutation invariant characteristics~\cite{adnan2020representation}.

\section{Method}\label{sec:method}

\begin{figure}[t]
  \centering
  \includegraphics[width=0.9\linewidth]{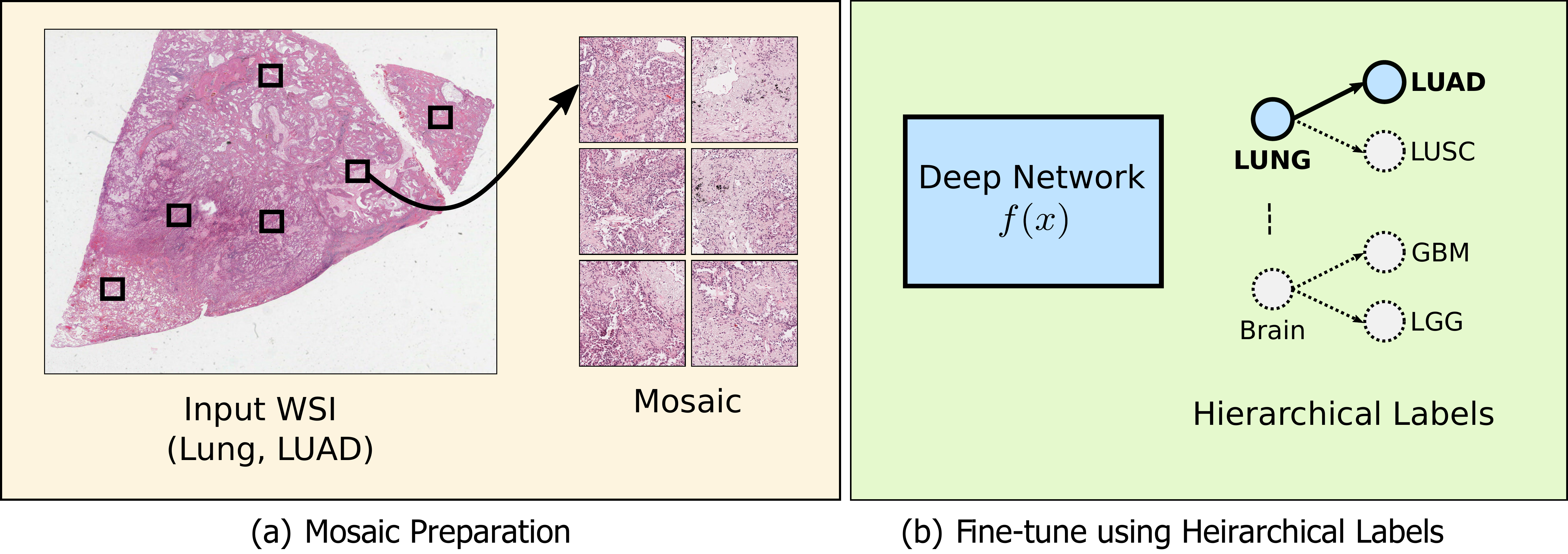} 
  \caption{\textbf{Training a Feature Extractor.} A feature extractor is trained with hierarchical target labels of a WSI. (a) A set of
    representative WSI patches (called mosaic) is
    extracted~\cite{kalra2020yottixel}. (b) The patches are used to fine-tune a deep
    network; each patch is
    assigned the parent WSI's labels, i.e., anatomic site and primary diagnosis.}
  \label{fig:stage1}
\end{figure}

\begin{figure}[t]
  \centering
  \includegraphics[width=0.95\linewidth]{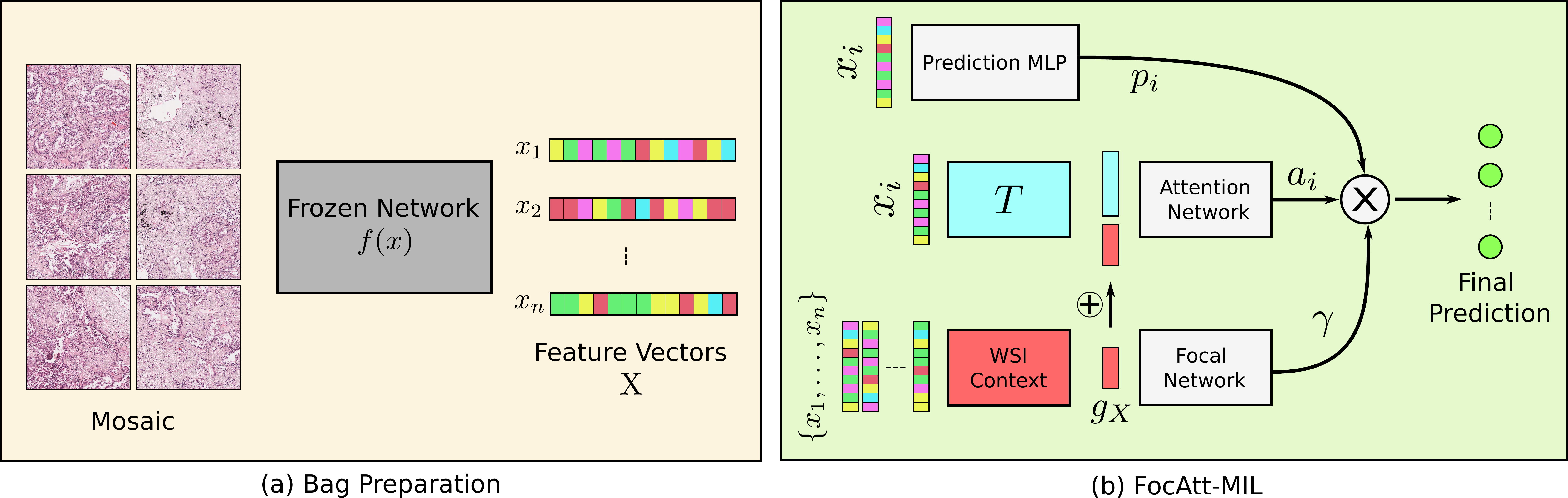} 
  \caption{\textbf{Classification of WSIs with FocAtt-MIL.} The two-stage method for the classification of WSI. (a) The mosaic of a WSI is
    converted to a bag $X$ containing a set of feature vectors
    $\{x_1,\dots,x_n\}$. (b) The feature vectors in a bag $X$ are transformed to the primary diagnosis probability through FocAtt-MIL. The prediction
    probability $p_i$ is computed for an individual feature vector $x_i$. A
    WSI context $g_X$  is computed for the entire bag $X$ using \eqref{eq:deepset}. The WSI context $g_X$ is used to compute the attention value
    $a_i$ and the focal factor $\gamma$. The final prediction is computed using~\eqref{eq:focatt}.}
  \label{fig:stage2}
\end{figure}

There are two stages in the proposed method (i)~bag preparation, and
(ii)~multi-instance learning with FocAtt-MIL. In the first stage, representative
patches (called mosaic) are extracted from a WSI. The mosaic's patches are
encoded to a set of feature vectors (called bag) using a deep network. The feature extraction model can be a pre-trained network, or can be fined-tuned to increase its effectiveness as shown in \autoref{fig:stage1}. In the second stage, the proposed MIL technique
(called FocAtt-MIL) is trained to predict the primary diagnosis for a given bag (a WSI). The schematic for the second stage is shown
in~\autoref{fig:stage2}.


\vspace{0.3cm}

\noindent \textbf{Bag Preparation.} A patch selection method proposed by Kalra
et al.~\cite{kalra2020yottixel} is used to extract the representative patches
from a WSI. We removed non-tissue regions using colour threshold. The
remaining tissue-containing patches are grouped into a pre-set number of
categories through a clustering algorithm. A portion of all clustered patches
(e.g., 10\%) are randomly selected within each cluster, yielding a \emph{mosaic}. The
mosaic is transformed into a bag $X=\{x_1,\dots,x_n\}$, where $x_i$ is the
feature vector of $i^{th}$ patch, obtained through a deep network (a  feature
extractor). The \autoref{fig:stage2} shows the bag preparation stage, the frozen
network $f(x)$ represents a non-trainable deep network used as a feature extractor.

\noindent \textbf{Fine-tune a Feature Extractor using Hierarchical Labels.} In
MIL, robust features enable weak learners to make better predictions thus
improving the final aggregated prediction. A WSI is generally associated with
the following two labels---anatomic site and primary diagnosis. These two
labels are arranged in hierarchy as shown in \autoref{fig:stage1}. Consider,
$y_{as}$ and $y_{pd}$ represent anatomic site and primary diagnosis
respectively. Then, instead of predicting these labels independently, we predict $P(y_{as})$, and $P(y_{pd}|y_{as})$. The conditional
probability $P(y_{pd}|y_{as})$ helps in modelling the dependent relationship.
Using Bayes theorem, we get, $ P(y_{as}|y_{pd}) =
{P(y_{pd}|y_{as})P(y_{as})}/{P(y_{pd})}$, where $P(y_{as}|y_{pd})=1$, because of
the dependence. We simplify $P(y_{pd}) = P(y_{pd}|y_{as})P(y_{as})$, and compute
cross entropy losses for the predictions of both $y_{as}$ and $y_{pd}$. We equally weight both the losses towards the final loss of the network.

\vspace{0.3cm}

\noindent \textbf{WSI Context Learning. } A single vector representation of a
WSI (or a bag $X$) is computed as, 
\begin{equation}
\label{eq:deepset}
g_X=\phi(\theta(x_1),\dots,\theta(x_n)),
\end{equation}
where, $\theta$ is a neural network and $\phi$ is a pooling function, such as
sum, mean, and max. It has been proven in~\cite{deepset} that \eqref{eq:deepset} can approximate any set function. The vector $g_X$
is used by the attention module and the focal network.

\vspace{0.3cm}

\noindent \textbf{The FocAtt-MIL Approach. } The FocAtt-MIL is a permutation-invariant
model that learns to predict a target label (primary diagnosis) $y_{pd}$ from a
bag $X$ (a WSI). The approach is composed of four major components (\autoref{fig:stage2}):

\begin{enumerate}
\item \textit{Prediction MLP.} A prediction $p_i$ is computed
  for each item $x_i$ in the bag $X$, using a
  trainable deep network called Prediction MLP.
\item \textit{WSI Context.} It a deep network that computes a single vector
  representing an entire bag $X$ using \eqref{eq:deepset}.
\item \textit{Attention Module.} The attention module is composed of two
  networks, a transformation network $T$, and the Attention Network. The attention module takes
  the $i^{th}$ patch $x_i \in X$,  and the WSI context $g_X$ to compute an attention
  value $a_i \in [0, 1]$ for that patch.
\item \textit{Focal Network.} Another deep network that uses WSI context $g_X$
  to compute a focal factor $\gamma$ (a vector) that modulates the final prediction. The length of $\gamma$ is
  same as the number of discrete values in the target label, thus allowing the
  per dimension modulation.
\end{enumerate}

\noindent \textbf{The Final Prediction.} The final output from the FocAtt-MIL is
computed by aggregating individual attention-weighted predictions modulated by
the learned focal factor, as follows
\begin{equation}
  \label{eq:focatt}
 y(j) = \sum_{i=1}^{n}{\mathbf{p_i}(j)^{\boldsymbol{\gamma}(j)} a_i}.
\end{equation}
The $\mathbf{p_i}$, and $\boldsymbol{\gamma}$ in \eqref{eq:focatt} are both vectors. The $y$ is converted to a probability distribution by dividing with $sum(y)$.

\section{Results}\label{sec:result}
We evaluated the proposed approach for two different WSI classification tasks.
All experiments are conducted with 4 Nvidia V100 GPUs (32 GB vRAM each). The
code has been written using the Tensorflow library~\cite{abadi2016tensorflow}.

\begin{table}[b]
  \centering
  \caption{Performance comparison for LUAD/LUSC classification via
    transfer learning.}
  \begin{tabular}{lc}
    \toprule
    \textbf{Algorithm}                              & \textbf{Accuracy}  \\
    \toprule
    Coudray et al.~\cite{coudray2018classification} & 0.85          \\
    Kalra \& Adnan et al.~\cite{kalra2020learning}  & 0.85          \\
    Khosravi et al.~\cite{khosravi2018deep}         & 0.83          \\
    Yu et al.~\cite{yu2016predicting}               & 0.75          \\
    \textbf{FocAtt-MIL (proposed method)}           & \textbf{0.88}\\
    \bottomrule
  \end{tabular}
  \label{tab:luad-lusc-res}
\end{table}

\vspace{0.3cm}

\noindent \textbf{LUAD vs LUSC Classification --} Lung Adenocarcinoma (LUAD) and
Lung Squamous Cell Carcinoma (LUSC) are two main subtypes of non-small cell lung
cancer (NSCLC) that account for 65-70\% of all lung cancers~\cite{zappa2016non}.
An automated classification of these two main sub-types of NSCLC is a crucial
step to assist pathologists~\cite{graham2018classification,zappa2016non}. For
this task, we establish the efficacy of FocAtt-MIL to differentiate between LUAD
and LUSC. We obtained 2,580 hematoxylin and eosin (H\&E) stained WSIs of lung
cancer from TCGA repository~\cite{tomczak2015cancer}. The data is split into
1,806 training, and 774 testing WSIs~\cite{kalra2020learning}. The dataset is
approximately 2 TB. We obtained mosaic for each WSI using the approach
in~\cite{kalra2020yottixel}, and subsequently converted the mosaic to a bag $X$
of features using a pre-trained DenseNet~\cite{huang2017densely}. We did not
fine-tune the feature extraction model for this task in order have a fair
comparison against other transfer-learning based approaches in the literature.
We trained the FocAtt-MIL to classify bags between the two sub-types of lung
cancer. We achieved the accuracy of 88\% on test WSIs (AUC of 0.92). The
accuracy has been reported in \autoref{tab:luad-lusc-res}.

We conducted an \textbf{ablation study} to understand the effect of different
model parameters. Removing the WSI context $g_X$ from the attention module,
resulted in 4\% reduction of the accuracy. Excluding the focal factor $\gamma$
and the global context $g_X$ from the final prediction, resulted in 6\%
reduction in the accuracy. The ablation suggests that the model's performance is
the most optimal by (i)~incorporating the WSI context $g_X$ in the attention
computation, and (ii)~allowing the focal factor to modulate the final aggregated
prediction.

\begin{figure}[t]
    \centering
    \includegraphics[width=0.45\linewidth]{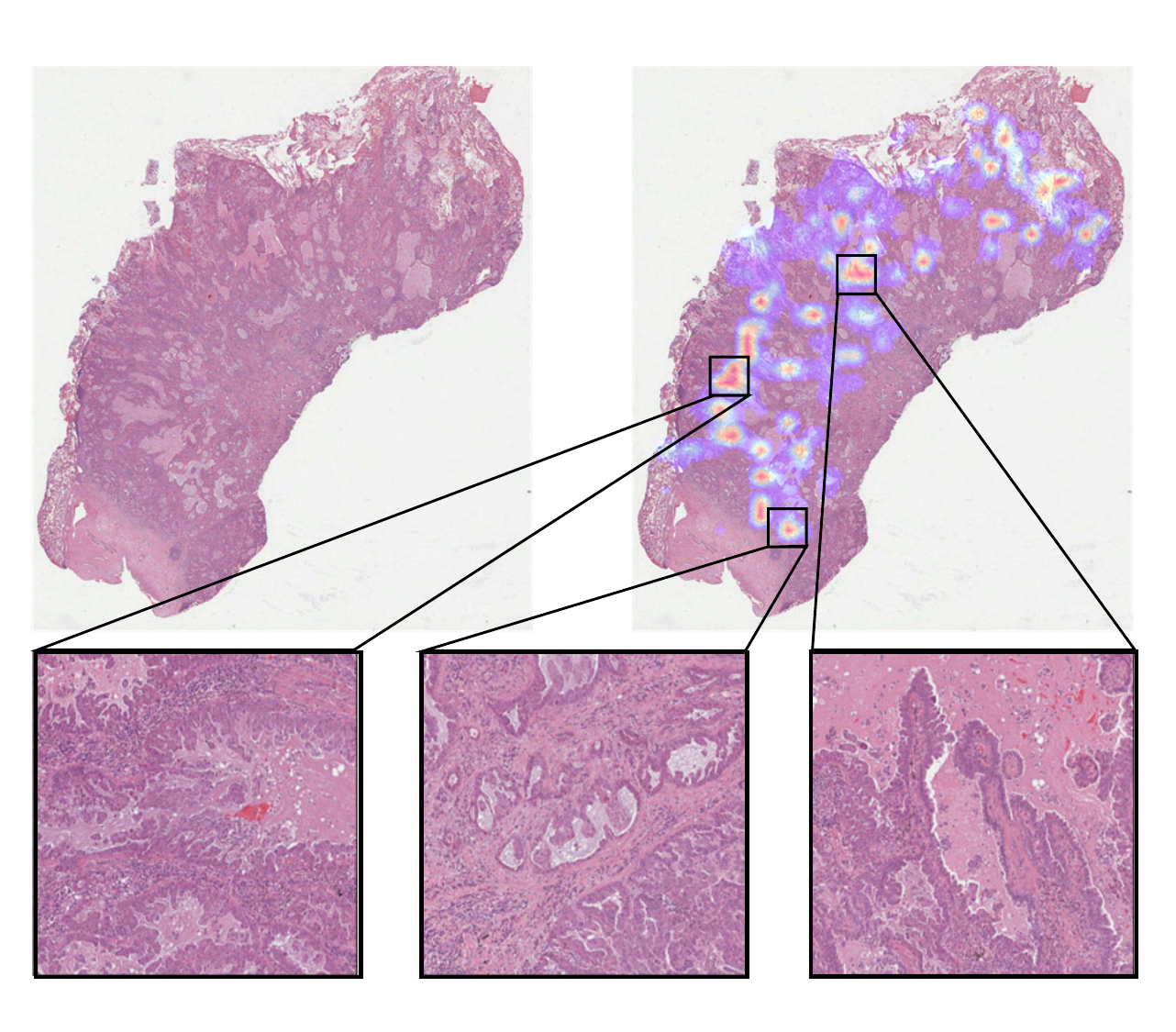}
  \centering
    \centering
    \includegraphics[width=0.43\linewidth]{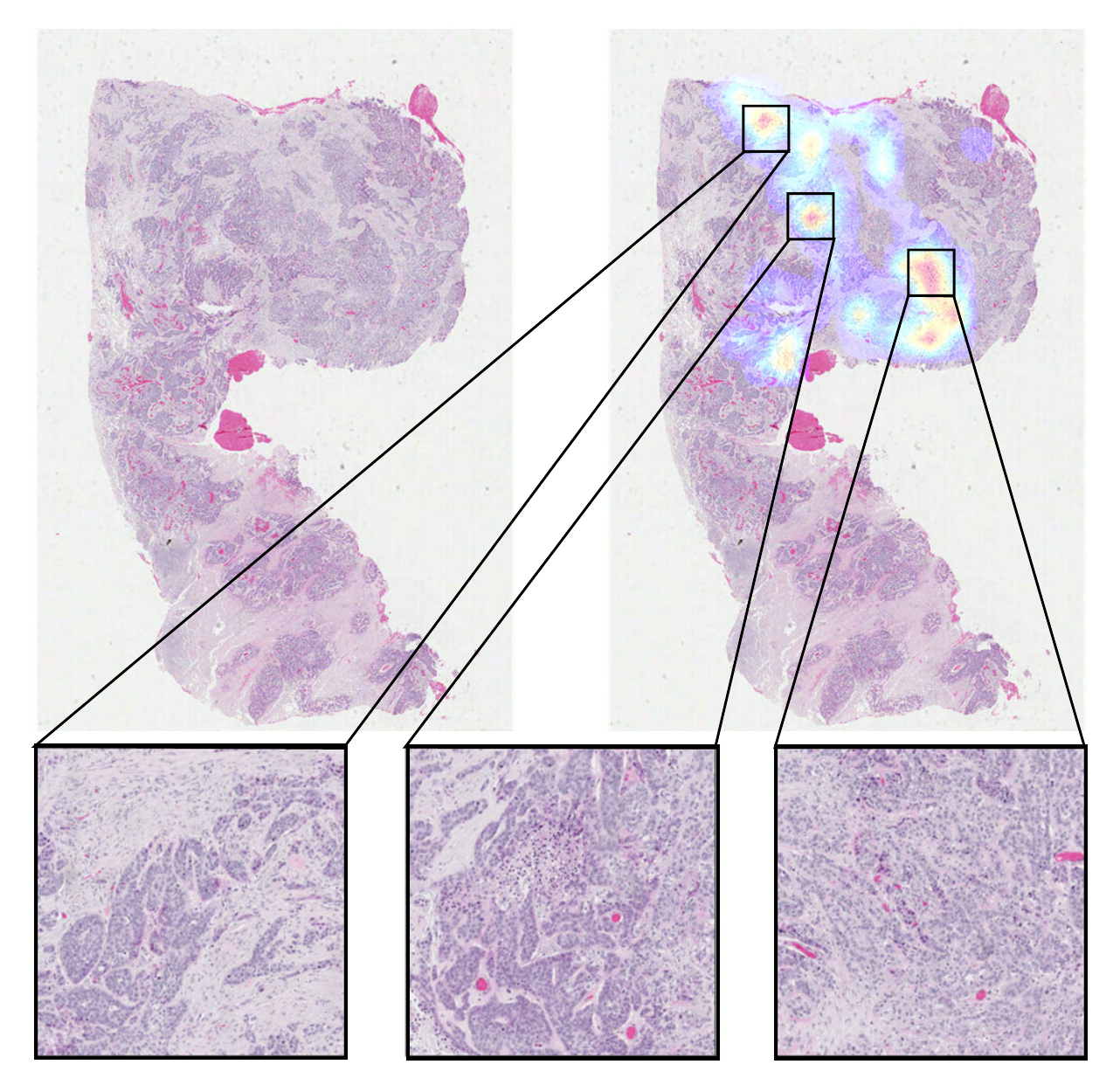}
    \caption{\textbf{Attention Visualization.} The attention values augmented
      on the two exemplar WSIs. \textbf{Left Image (LUAD):}~Regions of
      the highest importance come from the cancerous regions while sparing normal lung
      tissue, fibrosis, and mucin deposition. Additionally, by inspecting important
      regions at a higher magnification, it is noticeable that the malignant glandular
      formations border with non-malignant areas. \textbf{Right Image (LUSC):}~Regions that
      are considered to be important for classification are composed of malignant
      squamous cells. However, unlike LUAD, the attention model seems to be responsive
      to regions with solid malignant structures.}
  \label{fig:luad-lusc-attention}
\end{figure}

We used the attention module of the trained model to \textbf{visualize the
attention heat-map} on the unseen WSIs (\autoref{fig:luad-lusc-attention}). The
visual inspection of these two WSIs reveals that the model made its decision
based on regions containing malignant tissue and ignored non-cancerous regions.
In the LUSC WSI (right), regions with squamous formations are deemed the most
important ones. For the LUAD WSI (left), the salient regions are solely coming
from the malignant area, implying that the model differentiates between normal
lung alveolar tissue and LUAD. Therefore, one could say that attention heatmaps
are histopathologically meaningful. For LUAD samples, regions where cancerous
tissue meets non-cancerous structures are deemed the most important. Such
contrast makes cancerous glandular structures easier to recognize. However, this
phenomenon cannot be seen in LUSC samples, as the model is responsive to regions
that are completely composed of malignant squamous carcinoma.

\vspace{0.3cm}

\noindent \textbf{Pan-cancer Analysis -- } 
In the second experiment series, we evaluated the approach against a
large-scale pan-cancer classification of WSIs. The \textbf{dataset} used for this task has
been proposed by Riasatian et al.~\cite{riasatian2021fine}. It comprises more
than 7 TB data, consisting of 7,097 training, and 744 test WSIs, distributed
across 24 different anatomic sites, and 30 different primary diagnoses. All
WSIs in the dataset are taken from a public repository of WSIs,
TCGA~\cite{tomczak2015cancer}. We obtained a mosaic for each WSI, and then
applied a cellularity filter~\cite{riasatian2021fine} to further reduce the number of patches in each
mosaic. Subsequently, we obtained 242,202 patches for training WSIs and
116,088 patches for testing WSIs. Each patch is of the size 1000$\times$1000,
but we resized them to 256$\times$256 pixels.

We used three different \textbf{feature extractors} to validate the FocAtt-MIL.
We prepared a separate ``bag'' for each feature extractor. These three feature
extractors are: DenseNet (DN)~\cite{huang2017densely},
KimiaNet~\cite{riasatian2021fine}, and the fine-tuned DenseNet (FDN). We
fined-tuned the DenseNet on training patches using weak labels obtained from
their respective WSIs. The weakly labelled fine-tuning has shown to be
effective~\cite{riasatian2021fine}. In our case, the weak labels are anatomic
site, and primary diagnosis, arranged in a hierarchy. This hierarchical
arrangement of labels is incorporated during the training using the approach
outlined earlier in the Section~\ref{sec:method}. For the fine-tuning, we used
Adam optimizer~\cite{kingma2014adam} and a learning rate of $10^{-5}$ were used
for 20 epochs.

\begin{figure}[h]
  \subfigure[Validation Loss]{
    \centering
    \includegraphics[width=0.47\linewidth]{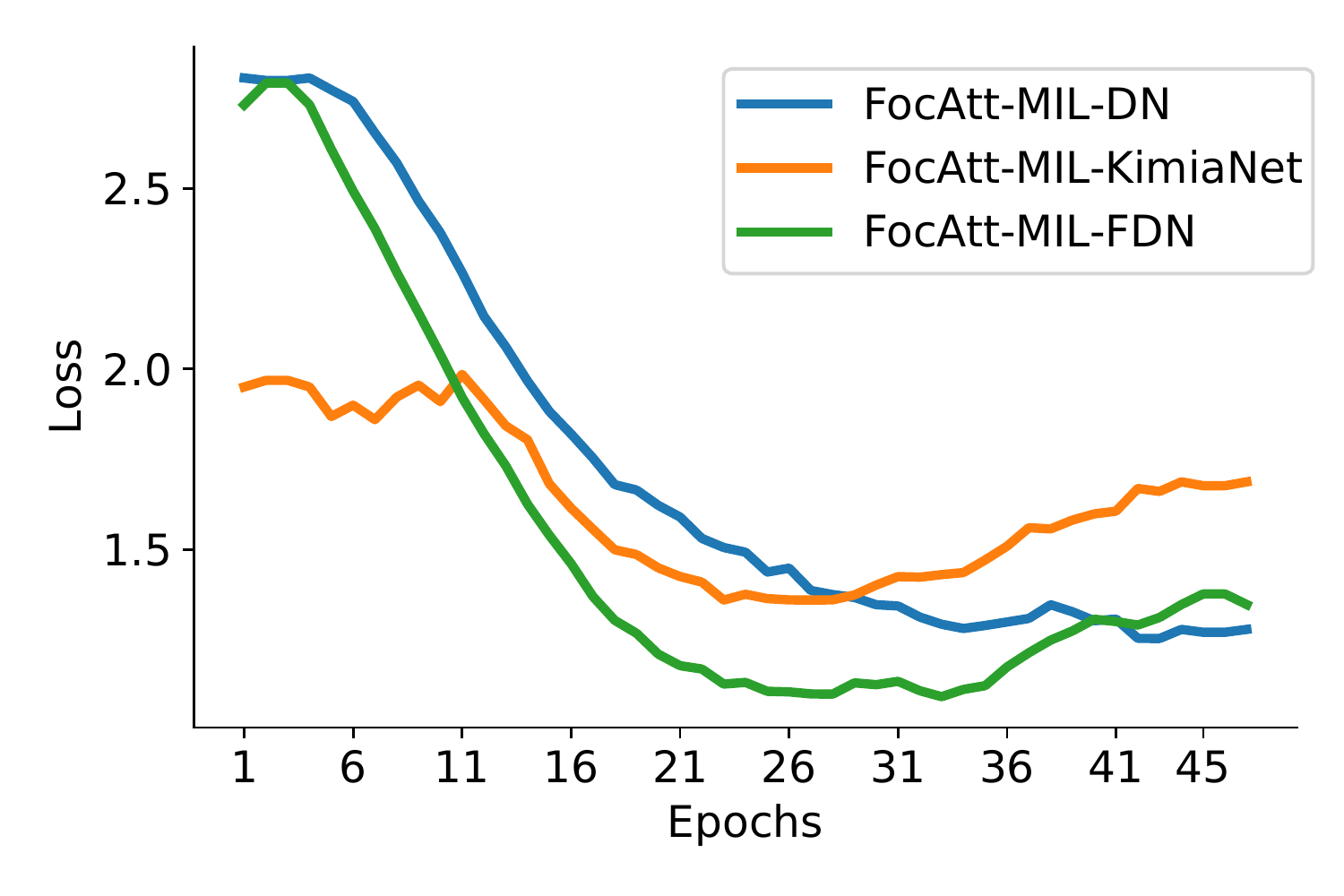}
  }
  \centering
  \subfigure[Validation Accuracy]{
    \centering
    \includegraphics[width=0.47\linewidth]{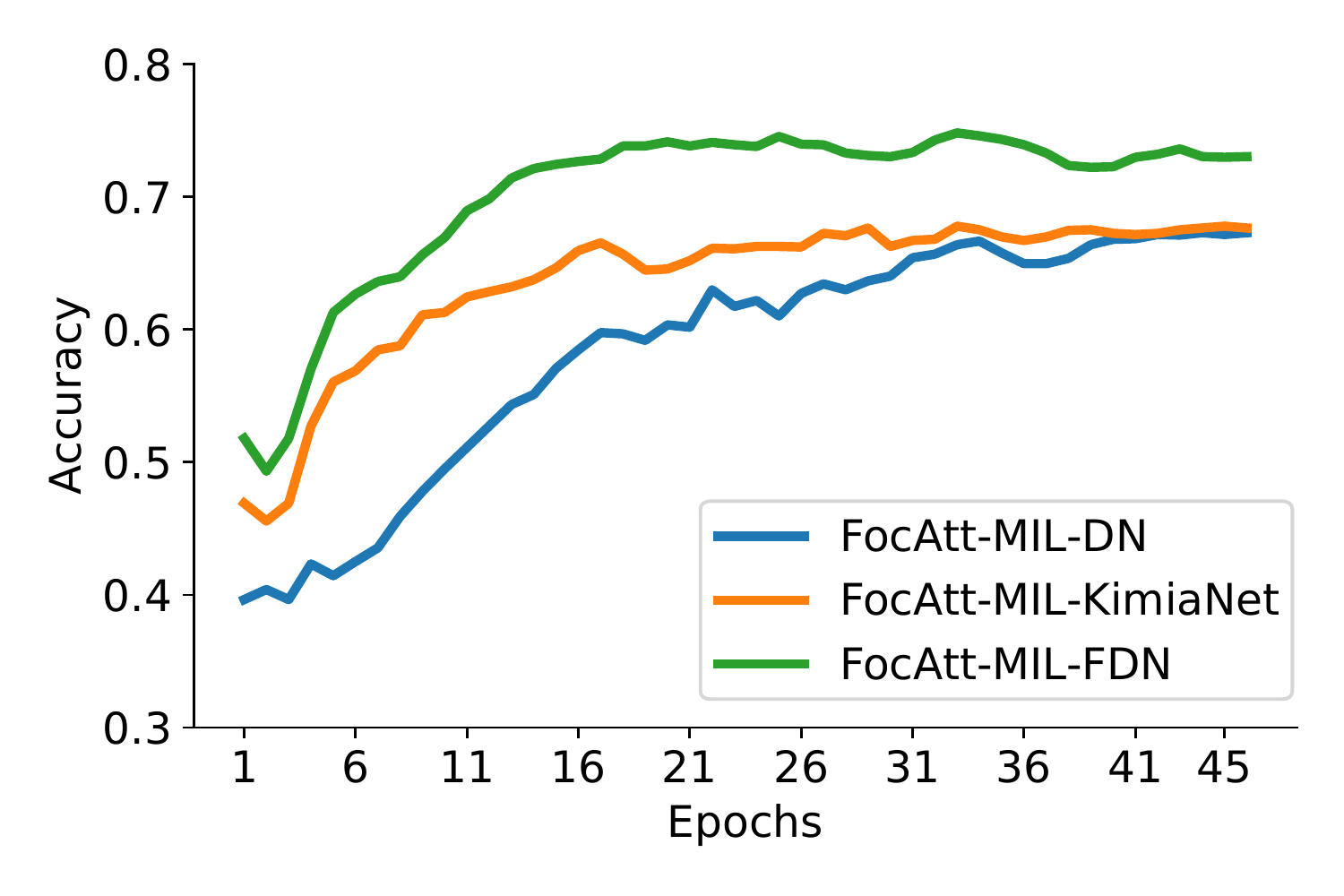}
  }
  \caption{\textbf{FocAtt-MIL Training.} The loss and accuracy on validation
    dataset during the training of three different configurations of FocAtt-MIL,
    i.e FocAtt-MIL-DN, FocAtt-MIL-KimiaNet, and FocAtt-MIL-FDN.
  }
  \label{fig:focatt-converge}
\end{figure}

We \textbf{trained the FocAtt-MIL} model with the same architecture for all the
three different bags. We tested three different configurations of FocAtt-MIL,
i.e., FocAtt-MIL-DN, FocAtt-MIL-KimiaNet, and FocAtt-MIL-FDN. For all the three
configurations, we used the SGD optimizer with a learning rate of 0.01, weight
decay of 10$^{-6}$, and momentum of 0.9. We applied \emph{gradient clipping} of
0.01 and dropout between layers to prevent the exploding gradients. We trained
models for 45 epochs. \autoref{fig:focatt-converge} shows the validation loss
and accuracy while training the three different configurations. It is evident
that FocAtt-MIL-FDN is outperforming from the very early epochs. It is
interesting to note that, both FocAtt-MIL-FDN, and FocAtt-MIL-KimiaNet (feature
extractors specialized for histopathology) seems to have converged to an optimal
validation accuracy around 20-25 epochs.

The 30 unique primary diagnoses in the dataset can be further grouped into 13
tumor types. The type of tumour is generally known at the inference time, and
the objective is to predict the cancer sub-type. To \textbf{validate the
efficacy of our model}, we computed the cancer sub-type classification (i.e.,
primary diagnosis) accuracy for the given tumour type. This type of
classification is called \emph{vertical classification}. The vertical
classification results are reported in
\autoref{tab:pan-cancer-result}\footnote{For abbreviations GBM, LGG, ACC,...,
see \emph{wiki.cancerimagingarchive.net}}. The results show that FocAtt-MIL can
elevate the accuracy of pre-trained features; DenseNet features have shown to
under-perform compared to KimiaNet
features~\cite{riasatian2021fine,kalra2020pan}. However, within the proposed
FocAtt-MIL scheme, DenseNet features become quite competitive. This applies to
the fine-tuned DenseNet (FocAtt-MIL-FDN) as well, whose results are on par with
the highly customized KimiaNet features when used within the FocAtt-MIL
framework.

\begin{table}[t]
  \centering
  \caption{Pan-cancer vertical classification accuracy of FocAtt-MIL for
    features from regular DenseNet (FocAtt-MIL-DN), KimiaNet (FocAtt-MIL-KimiaNet),
    and DenseNet fine-tuned with hierarchical labels (FocAtt-MIL-FDN).}
  \label{tab:pan-cancer-result}
    \scalebox{0.8}{ \begin{tabular}{p{3.8cm}p{3.2cm}m{2.1cm}m{2.4cm}m{2.3cm}}
\toprule
\textbf{Tumor Type}                                   & \textbf{Primary Diagnosis} & \textbf{FocatAtt-MIL-DN}      & \textbf{FocAtt-MIL-KimiaNet}        & \textbf{FocAtt-MIL-FDN}   \\
\toprule
  \multirow{2}{*}{Brain}                     & GBM               & \textbf{0.9714} & 0.9429          & 0.8571          \\
                                             & LGG               & 0.6410          & 0.7692          & \textbf{0.8205} \\ \hline
  \multirow{3}{*}{Endocrine}                 & ACC               & 0.6667          & 0.6667          & 0.6667          \\
                                             & PCPG              & \textbf{1.0000} & \textbf{1.0000} & \textbf{1.0000} \\
                                             & THCA              & 0.9608          & \textbf{1.0000} & \textbf{1.0000} \\ \hline
  \multirow{4}{*}{Gastrointestinal tract}    & COAD              & \textbf{0.6875} & 0.4375          & 0.5000          \\
                                             & ESCA              & 0.5000          & \textbf{0.8571} & 0.5714          \\
                                             & READ              & 0.0833          & 0.5000          & \textbf{0.6667} \\
                                             & STAD              & \textbf{0.8333} & 0.7333          & \textbf{0.8333} \\ \hline
  \multirow{3}{*}{Gynaecological}            & CESC              & 0.8824          & \textbf{0.9412} & 0.7647          \\
                                             & OV                & 0.5000          & 0.8000          & \textbf{1.0000} \\
                                             & UCS               & 0.6667          & \textbf{1.0000} & 0.3333          \\ \hline
  \multirow{3}{*}{Liver, pancreaticobiliary} & CHOL              & 0.2500          & 0.0000          & \textbf{0.5000} \\
                                             & LIHC              & 0.8857          & \textbf{0.9143} & 0.8571          \\
                                             & PAAD              & \textbf{1.0000} & 0.7500          & 0.8333          \\ \hline
  \multirow{2}{*}{Melanocytic malignancies}  & SKCM              & \textbf{0.9167} & 0.8750          & \textbf{0.9167} \\
                                             & UVM               & \textbf{1.0000} & 0.2500          & \textbf{1.0000} \\ \hline
  \multirow{2}{*}{Prostate/testis}           & PRAD              & \textbf{1.0000} & 0.9500          & \textbf{1.0000} \\
                                             & TGCT              & \textbf{1.0000} & \textbf{1.0000} & \textbf{1.0000} \\ \hline
  \multirow{3}{*}{Pulmonary}                 & LUAD              & 0.5789          & 0.8158          & \textbf{0.8947} \\
                                             & LUSC              & \textbf{0.9302} & 0.6977          & 0.7442          \\
                                             & MESO              & 0.6000          & \textbf{1.0000} & \textbf{1.0000} \\ \hline
  \multirow{4}{*}{Urinary tract}             & BLCA              & 0.9118          & \textbf{1.0000} & 0.8529          \\
                                             & KICH              & 0.5455          & 0.6364          & \textbf{0.7273} \\
                                             & KIRC              & 0.9200          & 0.9000          & \textbf{0.9600} \\
                                             & KIRP              & 0.5714          & 0.6786          & \textbf{0.7143} \\
\bottomrule
\end{tabular} }
\end{table}

\vspace{0.3cm}

\noindent \textbf{Conclusions --} The accelerated adoption of digital pathology
offers a historic opportunity to find novel solutions for major challenges in
diagnostic histopathology. In this study, we proposed a
novel attention-based MIL technique for the classification of WSIs. We introduced a focal factor, computed using a global representation
of WSI for modulating the individual patch-level prediction, thus promoting
more accurate aggregated final prediction. We also proposed a novel fine-tuning
approach to extract more robust features from WSI patches. We fine-tune a feature extraction model using patches and the hierarchical weak labels from
their respective WSIs. We validated the proposed framework on two large datasets
derived from TCGA repository~\cite{tomczak2015cancer}. The results suggest
competitive performance on both the datasets. Furthermore, the proposed method is
explainable and transparent as we utilized the attention values to visualize
important regions.

\bibliography{ref}   
\bibliographystyle{splncs04}

\end{document}